\documentclass[manuscript]{aastex}

\shorttitle{Crossed log-periodic dipole}
\shortauthors{Sasikumar et al.}

\begin{document}


\title{Design and performance of a low frequency cross-polarized log-periodic dipole antenna}


\author{K. Sasikumar Raja, C. Kathiravan, R. Ramesh, M. Rajalingam and Indrajit V. Barve}
\affil{Indian Institute of Astrophysics, II Block, Koramangala, Bangalore -560 034.\\}
\email{sasikumar@iiap.res.in}

%\altaffiltext{}{Indian Institute of Astrophysics, II Block, Koramangala, Bangalore -560 034.}


\begin{abstract}
We report the design and performance of a cross-polarized log-periodic dipole
(CLPD) antenna for observations of polarized radio emission from the
solar corona at low frequencies. The measured isolation between the
two mutually orthogonal log periodic dipole antennas was as low as
$\approx$ -43 dBm in the 65-95 MHz range. We carried out 
observations of the solar corona at 80 MHz with the above CLPD and 
successfully recorded circularly polarized emission. 
\end{abstract}



%% In the first two sections, notice the use of the natbib \citepp
%% and \citept commands to identify citations.  The citations are


\section{Introduction}

Circularly polarized radio radiation in the VHF 
range (30-300 MHz) can be 
received / transmitted using helical antennas, conical log-spiral antennas,
cross-polarized Yagi-Uda 
antennas, cross-polarized log-periodic dipole antennas, etc. In situations
where wider frequency coverage (10:1 or even more) is required, 
cross-polarized log-periodic dipole (CLPD) antennas are generally
used. A CLPD consists of two linearly-polarized log-periodic dipole (LPD) 
antennas \citep{Ham57} fixed to a common axis in a mutually orthogonal fashion. 
Log-periodic antennas are widely used in the field of radio 
astronomy, particularly where simultaneous multi-frequency observations of radio
emission from the celestial radio sources are required \citep{Eri74,Boi80,Maa13}. 
For example, in the case of the solar corona, radio emission at 
different frequencies orginate at different levels in the atmosphere.
To obtain data on the activities related to a solar flare, which leads
to the generation of transient radio emission almost around the same time 
at different levels in the solar corona, simultaneous multi-frequency 
observations are required \citep{Ram98}. 
However, the typical isolation between the two mutually orthogonal LPDs in commercially 
available CLPDs is less, $\approx$ -20 dB \citep{Piv06}. Because
of this limit, an understanding of the 
polarization characteristics of the weak signals from
celestial radio sources can be limited. Note that  
a $90^{o}$ hybrid is generally used 
in conjunction with the CLPD and the polarization strength is measured 
from the difference of the two outputs of the CLPD. The problem here is that the
outputs from either LPD in a CLPD responds to the total
intensity also. So the observer encounters the problem of measuring a small
difference between two much larger quantities similar to polarization
observations with circular feeds \citep{Ric07}.
Though dual-polarized antenna designs
offering improved isolation have been mentioned in the literature, they 
are primarily at frequencies $>$ 1 GHz \citep{Piv06,Tra07}. 
Our interest
is to observe the polarized radio emission from the solar corona 
with high accuracy at low frequencies and use it to estimate the solar 
coronal magnetic field, one of the holy grails in solar astrophysics.
Hence this paper is presented to achieve this goal. Note that low frequency radio emission 
originates from regions of the solar atmosphere where observations
in whitelight and other regions of the electromagnetic spectrum are 
presently difficult.

\section{Design and fabrication of the CLPD}

\begin{figure}[!t]
\centering
\includegraphics[width=4in]{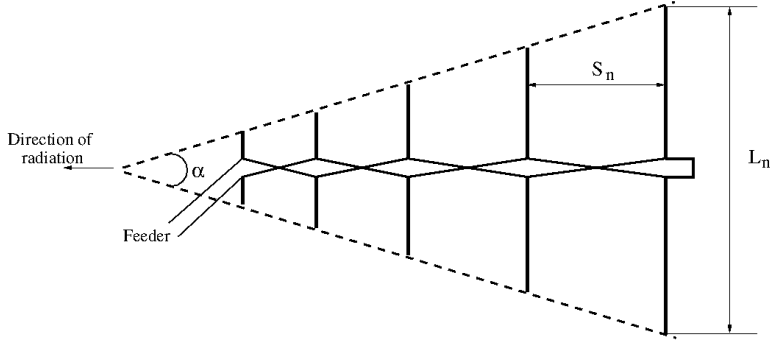}
\DeclareGraphicsExtensions.
\caption{Schematic design of a LPD.}
\label{fig1}
\end{figure}

A step-by-step procedure for designing a LPD
was first described by \citet{Car61}. In our
efforts to construct a CLPD, we used the
inputs mentioned in the above reference and fabricated a 
LPD. 
Figure \ref{fig1}
shows the schematic design of a LPD. It can be shown that
the apex angle $\alpha$ is related to the length ($L_{n}$) of 
the adjacent arms and the spacing ($S_{n}$) between them as
\citep{Kra50,Bal05}:
\begin{equation}
L_{n+1} = L_n + S_n~{\rm tan}\alpha \label{one}
\end{equation}
The length of the adjacent arms and the spacing between them follow
the relationship:
\begin{equation}
{L_{n+1}\over{L_n}}={{S_{n+1}\over{S_n}}}=k \label{two}
\end{equation}
where $k$ is a constant. The frequencies ($f$) at which a LPD has 
identical performance are related by the following equations: 
\begin{equation}
f_n=f_{n+1} \star k \label{three} \\
\end{equation}
\begin{equation}
log(f_{n+1})=log(f_n)+log(1/k) \label{four}
\end{equation}
\noindent Using equation \ref{one} and by fixing the value of
$\alpha$, the length of the adjacent arms 
and the spacing between them are calculated iteratively. 
By making use of the 
optimum design curve of a LPD \citep{Car61}, the directional gain ($G$) of the 
LPD can be decided as a function of apex angle ($\alpha$), inter-arm spacing ($S$) 
and the scale 
factor $k$. 
In the present case,  
parameters were chosen as: $G$ = 8 dBi, $\alpha$ = $21^o$,
$k$ = 1.14, $S_\lambda$ = $S/\lambda$ = 0.08. In practice by using the 
optimum gain curve, we can choose $k$ for a 
required gain or vice-versa. The relationship between the 
bandwidth or the frequency ratio of a LPD, i.e. $F=f_{max}/f_{min}$
(where $f_{max}$ and $f_{min}$ are the 
expected maximum and minimum operating frequencies of the LPD) and the 
design parameters is given by,\\
\begin{equation}
k^N = F  \nonumber\\
\end{equation}

\begin{equation}
N = {{log(F)}\over{log(k)}} \label{five}
\end{equation}
where $N$ is the number of dipoles.
Once the values of $k$ and $F$ are decided, the number of possible arms 
can be obtained using equation \ref{five}. 

Our interest was to design a CLPD for operation 
in the frequency range $\approx 65-95$ MHz 
since the existing radioheliograph at the Gauribidanur 
radio observatory (located about 100 km north of Bangalore in India),
where the present work was carried out, operates primarily
at 80 MHz \citep{Ram98,Ram99,Ram06,Ram11b}.
The number of arms generally 
vary with the frequency coverage of the antenna and in the present
case it was decided to fabrciate the LPD with 4 arms, based on the values 
of $F$ and $k$ 
mentioned above.
The lengths of the different arms of the LPD calculated
for the above frequency range are tabulated in Table \ref{table1}. 
The arms were designed using commerically available hollow cylindrical
aluminium pipes of diameter $\approx$ 13 mm and they were fixed (in pairs) 
to two identical 
hollow rectangular aluminium 
pipes (called the booms) separated from each other by a non-conducting 
spacer in a criss-cross fashion (see Figure \ref{fig1}). 
The inter-boom separation ($D$) was 
calculated using the following equation \citep{Wak99}: 
\begin{equation}
Z_o = 138~log_{10}\left({{2 \sqrt{2} D} \over d}\right)  \label{six}
\end{equation}
where $Z_{o}$ is the characteristic impedance of the LPD and $d$ is the width of 
each boom. 
We used the commercially available hollow rectangular pipes with $d \approx 2.5$ cm
for the boom. We considered $Z_{o}=50~\Omega$ since the LPD output
was tapped using a RF coaxial cable as mentioned below whose 
characteristic impedance is $50~\Omega$.
Substituting the above values in equation \ref{six},
we found that $D \approx$ 2 cm.
The two booms act like a 2-conductor transmission line. They 
were `shorted' at one of their ends, close to where
the arms with the longest length are fixed. The distance
between the latter and the `short' (stub) is $\approx$ 29 cm.
This is $\rm {1/4}^{th}$ of the length $\rm L_{n}$ of the longest arm
in the CLPD, i.e. quarter wavelength loop (see Table \ref{table1}).
The output was tapped 
using a RF coaxial cable connected to the 
other end of the two booms (close to where the arms with the shortest 
length are fixed). While the center conductor of the RF coaxial cable 
was connected to one of the booms, the shield (i.e. the `ground') of the cable was
connected to the other boom and the coaxial cable was drawn through the latter
as described in \citet{Car61}.
The `shorting' of the two booms minimizes the impedance mismatch that 
arises when directly connecting the `unbalanced' RF coaxial
cable to the `balanced' LPD. The approximately quarter wavelength
of the coaxial cable in the boom acts as a `balun' by presenting a high
impedance to any common mode current \citep{Kra50,Bal05}.
We fabricated two LPDs with the above specifications 
and combined them in a mutually orthogonal fashion to form a CLPD. 
We measured the isolation of the CLPD as discussed in Section III and found that
it is about $\approx$ -20 dBm, nearly the same as that of a commercial CLPD
mentioned earlier.

\begin{figure}[!t]
\centering
\includegraphics[width=4in]{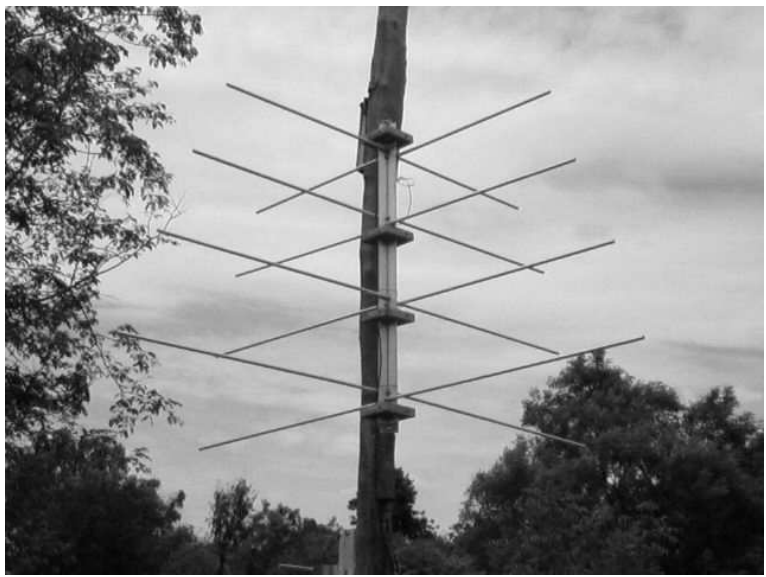}
\DeclareGraphicsExtensions.
\caption{A view of the low frequency CLPD designed and fabricated 
at the Gauribidanur radio observatory.}
\label{fig2}
\end{figure}
\begin{table}
\renewcommand{\arraystretch}{1.3}
% if using array.sty, it might be a good idea to tweak the value of
% \extrarowheight as needed to properly center the text within the cells
\caption{Antenna Specifications}
\label{table1}
\begin{center}
%\centering
% Some packages, such as MDW tools, offer better commands for making tables
%% than the plain LaTeX2e tabular which is used here.
\begin{tabular}{cccc}
\hline
S.No & $\rm L_{n}$ & $\rm S_{n}$ & Frequency \\
& (cm) & (cm) & (MHz) \\

\hline
1 & 79 & - & 95 \\
2 & 90 & 29 & 83 \\
3 & 102 & 31 & 74 \\
4 & 116 & 36 & 65 \\
\hline
\end{tabular}
\end{center}
\end{table}
It has been suggested that the isolation can be improved by decreasing the inter-boom 
spacing $D$ to $\approx 1/100^{th}$ of the arm length \citep{Piv06}. 
So, we designed a LPD/CLPD with $D \approx$ 5 mm
(smaller than even $1/100^{th}$ of the shortest arm length in Table \ref{table1}).
Using commercially available rectangular aluminium flats as the booms
instead of the rectangular hollow aluminimum pipes mentioned above. All the
other specifications including the characteristic impedance were the same.
The new CLPD mentioned above 
is shown in Figure \ref{fig2}. The RF cable 
was enlcosed inside a small aluminimum pipe fixed by the side of the 
aluminium flat that was `grounded'.
The isolation was measured to be $<$ -30 dBm. The details of the 
isolation and the other measurments carried out with the above CLPD are 
described in Section III. Trial observations with the CLPD are mentioned
in Section IV.

\section {Measurement of the VSWR, isolation and radiation pattern of the CLPD}
\subsection{VSWR measurements}
\begin{figure}[!t]
\centering
%\clearpage
\includegraphics[width=4in]{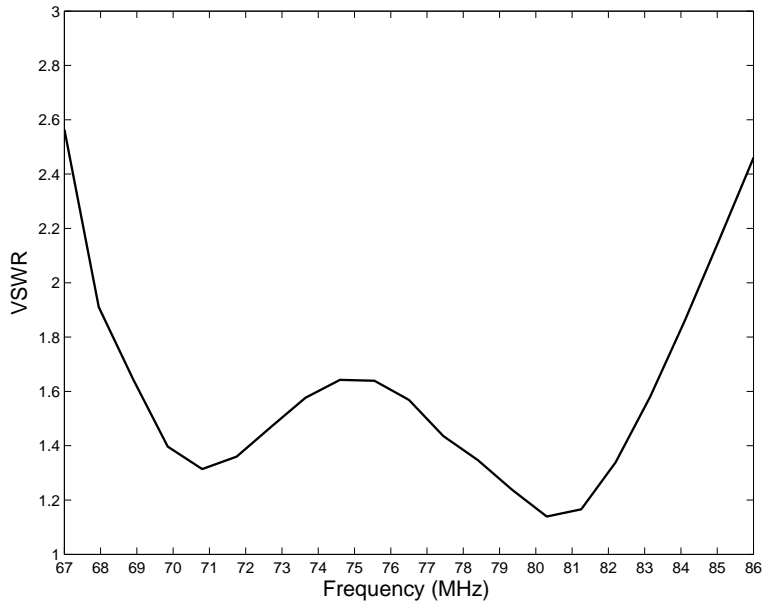}
\DeclareGraphicsExtensions.
\caption{VSWR of the CLPD in Figure \ref{fig2}.}
\label{fig3}
\end{figure}

The Voltage Standing Wave Ratio (VSWR) of the LPD/CLPD mentioned above 
was measured using a vector network analyzer. 
The values were found to be $\lesssim2$ for both the LPDs
in the frequency range of $\approx 68-84$ MHz (see Figure \ref{fig3}). The VSWR remained
nearly the same even after combining the 
two LPDs to form a CLPD. Due to the radio frequency interference generated by
FM transmissions ($88-108$ MHz) the higher frequency cut-off in our 
VWSR measurements were limited to $\approx$ 84 MHz.

\subsection{Far field radiation pattern measurements}

\begin{figure}[!t]
\centering
%\clearpage
\includegraphics[width=4in]{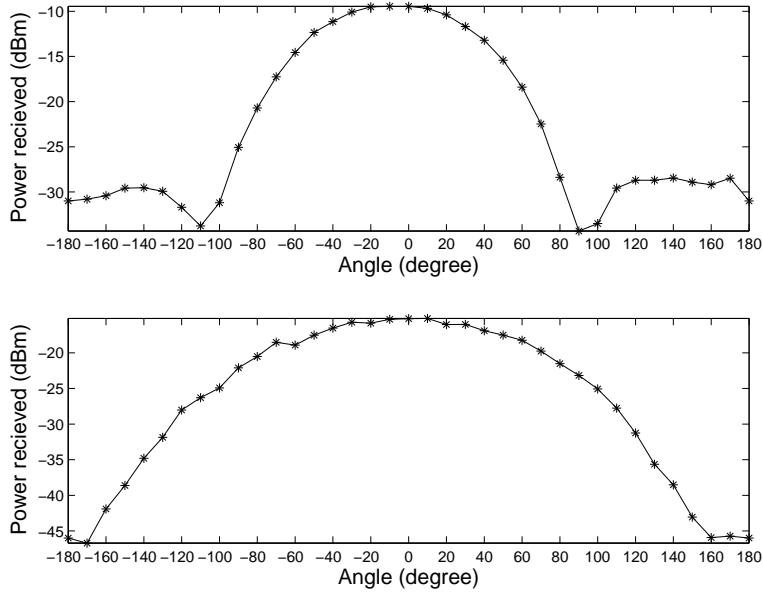}
\DeclareGraphicsExtensions.
\caption{Far field pattern of the CLPD 
in Figure \ref{fig2} at 80 MHz. The upper and lower panels correspond to the E-plane and H-plane, respectively.}
\label{fig4}
\end{figure}

To measure the far field radiation (Fraunhofer region) pattern, the receiver antenna and the transmitter antenna
were mounted at the height of $\approx$ 3 m above the ground level. The antennas
were separated by a distance of $\approx$ 3 m consistent with the
theoretical minimum observation distance for far-field measurements, i.e. $r_{ff} \gtrsim 2 l^2/\lambda$,
where $l$ is the length of the longest dipole in the CLPD and $\lambda$ is the
wavelength of the transmitted signal \citep{Bal05}. At 80 MHz,
$r_{ff} \approx$ 3 m for $l \approx 2.32$ m (see Table \ref{table1}).
The signal was transmitted using a CLPD and the same was received by the arms of 
a LPD in the same plane. 
The measurements
were carried out for different angle by rotating the transmitting
LPD in the azimuth direction. This corresponds to the E-plane measurement.
By mounting the transmitting and the receiving antennas in the vertical 
direction the measurements were repeated.
This corresponds to the H-plane measurement. 
Figure \ref{fig4}
shows the E-plane and H-plane far field patterns at 80 MHz for the CLPD in Figure \ref{fig2}.
The half power beam width (HPBW) in the 
E-plane and H-plane are $\theta \approx 60^{o}$ and $\phi \approx 120^{o}$, respectively.
We repeated the test at different frequencies and found that the E-plane and H-plane widths
are similar. The above beam widths correspond to a solid angle, $\Omega=\theta\phi \approx 2.2$ sr. 
From this we calculated
the directional gain of the antenna with respect to an isotropic radiator, i.e.
$G=10 {\rm log_{10}}(4\pi/\Omega) \approx$ 7.6 dBi. The effective collecting area is
$A_{e}=(G/4\pi)\lambda^{2}\approx 0.6 \lambda^{2}$.
The various parameters of the CLPD are listed in Table \ref{table2}. One can notice
a little asymmetry in the E-plane and H-plane far field patterns in Figure \ref{fig4}. 
The presence of residual common mode currents could be a reason for this. They may be
rejected to a large extent by adding clamp-on ferrite chokes to the coaxial cable. 
We plan to report this after the commissioning of a larger array with improved version
of the CLPD described in the present work.

\begin{table}[!t]
\begin{center}
\renewcommand{\arraystretch}{1.3}
% if using array.sty, it might be a good idea to tweak the value of
% \extrarowheight as needed to properly center the text within the cells
\caption{Calculated parameters of the CLPD in Figure \ref{fig2}}
\label{table2}
%\centering
% Some packages, such as MDW tools, offer better commands for making tables
%% than the plain LaTeX2e tabular which is used here.
\begin{tabular}{l l l}
\hline
S.No & Parameter & Value  \\

\hline
1 &HPBW in the E-plane ($\theta$) & $\approx 60^{o}$  \\
2&HPBW in the H-plane ($\phi$) & $\approx 120^{o}$ \\
3&Solid angle ($\Omega$) & $\approx 2.2$ sr\\
4&Directional gain ($G$) & $\approx 7.6$ dBi\\
5&Effective collecting area ($A_{e}$) & $\approx 0.6 \lambda^2$\\
\hline
\end{tabular}
\end{center}
\end{table}

\subsection{Near field radiation pattern measurements}
\begin{figure}[!t]
\centering
%\clearpage
\includegraphics[width=4in]{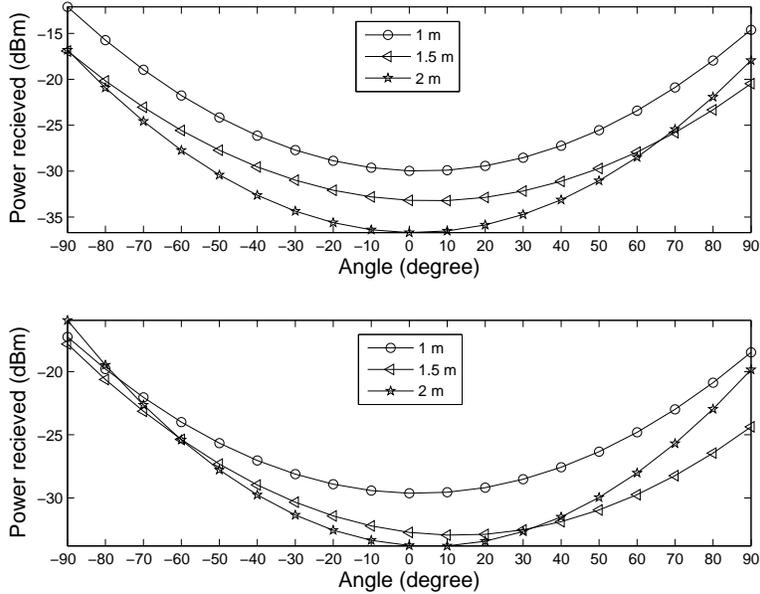}
\DeclareGraphicsExtensions.
\caption{Radiating near field pattern of the CLPD 
in Figure \ref{fig2} at 80 MHz. The upper and lower panels correspond to the E-plane and H-plane, respectively.
The numbers 1, 1.5 and 2 m in the rectangular box indicate the distances from the apex of the CLPD
at which the measurements were obtained.}
\label{fig5}
\end{figure}

To obtain the radiating near field (Fresnel region) pattern, we mounted the CLPD on a pole at a height 
of $\approx$ 3 m above the ground level and CW signal was transmitted from one of the LPDs in the
CLPD. Using a monopole antenna the signal strength was measured at different distances in the
radiating near field range, i.e. $0.62 \sqrt{l^{3}/\lambda} \lesssim r_{nf} \lesssim 2 l^{2}/\lambda$, from the apex
of the CLPD 
and also for different azimuth angles in the range $0^{o}-180^{o}$. 
At 80 MHz,  the radiating near field is in the distance range 
$1~{\rm m} \lesssim r_{nf} \lesssim 3$ m for $l \approx 2.32$ m (see Table \ref{table1}). 
The test was repeated for
the same set of distances and angles by transmitting
the signal through the other orthogonal LPD. The results were similar.
Figure \ref{fig5}
shows the E-plane and H-plane radiating near field patterns at 80 MHz for the CLPD 
in Figure \ref{fig2}. One can notice that the measured power varies with the distance
in the radiating near field range as expected.

\subsection{Isolation measurements}

\begin{figure}[!t]
\centering
%\clearpage
\includegraphics[width=4in]{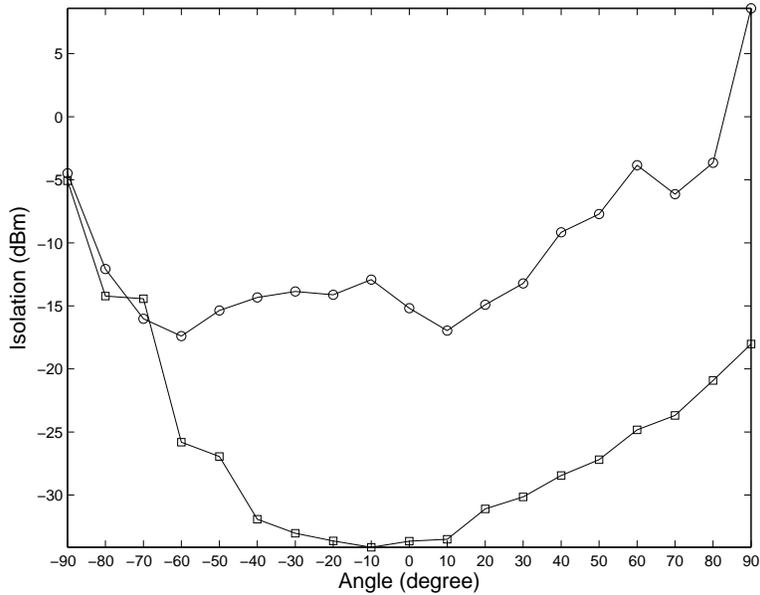}
\DeclareGraphicsExtensions.
\caption{Isolation between the orthogonal LPDs in the CLPD with inter-boom 
spacing $D$ = 5 mm (`squares') and $D$ = 2 cm (`circles') at 80 MHz.}
\label{fig6}
\end{figure}

\begin{figure}[!t]
\centering
%\clearpage
\includegraphics[width=4in]{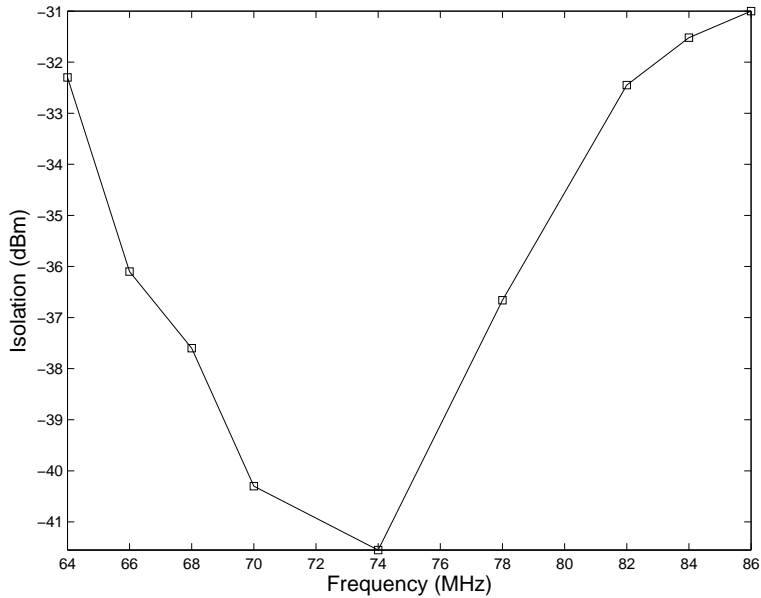}
\DeclareGraphicsExtensions.
\caption{Measured values of isolation (for an azimuthal angle of $0^{o}$) 
at different frequencies for the CLPD in Figure \ref{fig2} (inter-boom spacing $D$ = 5 mm).}
\label{fig7}
\end{figure}

By using the same set-up used to measure the far field pattern, we estimated the 
cross-talk and isolation in the CLPD. The transmitting LPD was mounted in the horizontal 
orientation ($0^{o}$) and the signal was received 
by both $0^{o}$ and $90^{o}$ oriented arms of the CLPD. In principle, we should receive signal 
in only the $0^{o}$ oriented arms of the CLPD. However, a finite signal was received 
in $90^o$ oriented arms of the CLPD. This is due to the cross-talk between the two
orthogonal LPDs in the CLPD. The difference in the signal strength received by the two orthognal
LPDs in the CLPD is a measure of the isolation. The measurements were repeated for different azimuth angles
and the results are shown in Figure \ref{fig6} for a frequency of 80 MHz. The two plots correspond to the
CLPD designed with inter-boom spacing ($D$) of $\approx$ 2 cm and $\approx$ 5 mm mentioned earlier. While the
isolation is $\approx $ -15 dBm for the former, it is $<$ -30 dBm for the latter. 
The variation of isolation with frequency is shown in Figure \ref{fig7}. 
We also estimated the isolation in an independant manner by transmitting the CW signal through one of the LPDs in the
CLPD in Figure \ref{fig2} and measuring the received power near the other LPD using a monopole. The results 
obtained were similar to that in Figure \ref{fig6}. We would like to note here that the 
isolation bandwidth depends on the extent to which the residual common mode currents (see Section 3.2) 
are rejected. It is possible that the limited bandwidth in Figure \ref{fig7} could be
due to the presence of such weak currents in our system.

\section{Observations}

\begin{figure}[!t]
\centering
%\clearpage
\includegraphics[width=4in]{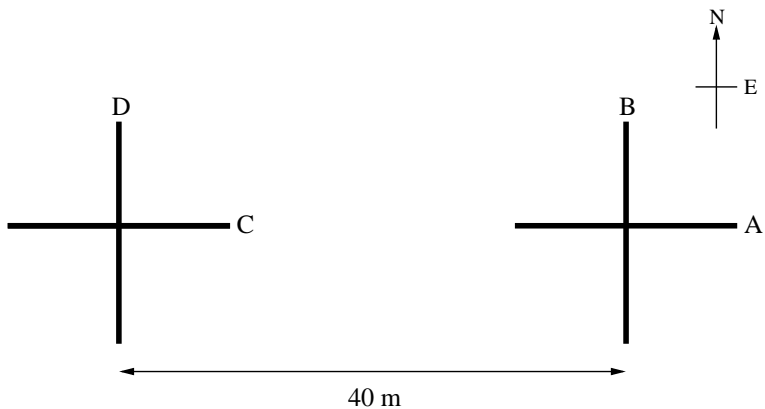}
\DeclareGraphicsExtensions.
\caption{Schematic of the interferometric polarimeter set up using the CLPDs
at the Gauribidanur observatory.}
\label{fig8}
\end{figure}

Two CLPDs similar to Figure \ref{fig2} were designed and mounted with a 
separation of $\approx$ 40 m in the
East-West direction. The CLPDs were at a height of $\approx$ 3 m above the
ground. They were mounted vertically in such
a manner that the length
of the arms gradually increase from the top to the base. 
The schematic 
diagram of the antenna set-up is shown in Figure \ref{fig8}. RF output
from the LPDs A, B, C and D 
were transmitted to a receiver room (about 500 m away) via low loss coaxial transmission 
lines buried under the
ground at a depth of $\approx$ 1 m to minimize the phase variations. 
We operated the
set-up as a correlation interferometer in the transit mode. The interferometer
technique has the advantage of: 1) minimizing the contribution from the
galactic background radiation and thereby the
emission from discrete sources can be observed with better contrast;
2) spurious effects due to ground
radiation are less \citep{Mor64}; and 3) calibration of the observations
is also relatively simpler \citep{Wei73,Sau96,Ram08}. In Figure \ref{fig8}, the multiplications
A x C and B x D respond to the total intensity (Stokes I) and the multiplications 
B x C and A x D respond to the circularly polarized intensity (Stokes V). In
principle the multiplications A x C and B x D record only 50\% of the total intensity.
Since the situation is the same for observations on the target as well as the 
calibrator sources, the error will be minimal.
Note that our interest is primarily on observations of the 
Stokes V emission from the solar corona since 
Stokes Q and U that contain information on the linear polarization of the signal, 
are considered to be extremely small at frequencies $<$ 100 MHz, particularly over
observing bandwidths $\gtrsim$ 1 MHz. The Faraday rotation of 
the plane of linear
polarization (during transmission through solar corona and the Earth's ionosphere) 
is considered to cancel the linear polarization generated at the source when the emission 
is summed over the observing band \citep{Gro73}. There are reports of observations
of high levels of linearly polarized radio emission from the Sun at 
frequencies $<$ 100 MHz, specifically
over narrow bandwidths in the range $\approx$ 0.1-10 kHz
\citep{Bho64,Chi71}. This needs to be verified. 
We would like to add here the interferometer method 
of measuring the Stokes V intensity described above differs from the 
conventional technique where a four-port $90^{o}$ hybrid is used in conjunction with
a crossed dipole feed to extract the Stokes V from the difference of the two outputs
from the hybrid. This method is also less sensitive to the cross-talk that arises
in a hybrid \citep{Coh58}.
The analog and digital correlator receiver system used for the observations are similar to that
described in \citet{Ram98}.
 
\begin{figure}[!t]
\centering
\includegraphics[width=4in]{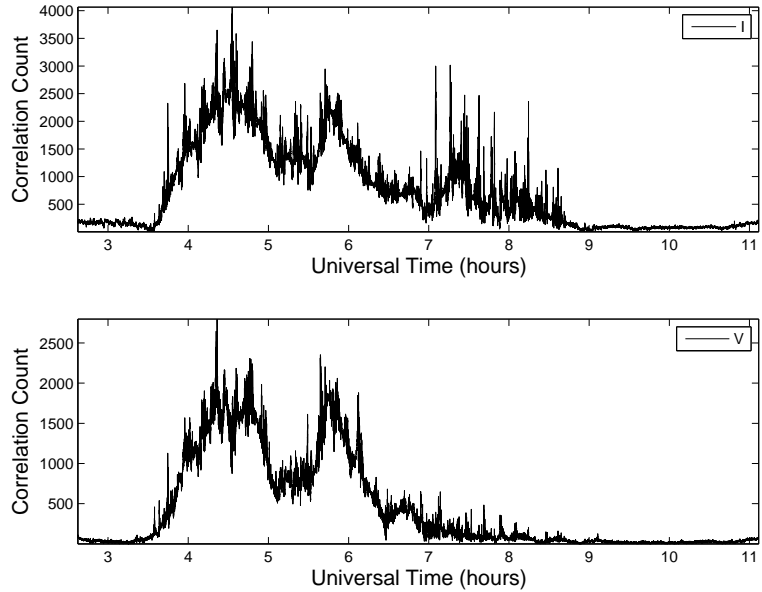}
\DeclareGraphicsExtensions.
\caption{Stokes I and V emission observed from the solar corona at 80 MHz on 
5 October 2012. The top and the lower panel corresponds to Stokes I and V
observations, respectively.}
\label{fig9}
\end{figure}

We carried out observations of the Sun and a few other unpolarized strong cosmic radio 
sources (calibrator sources) 
in both Stokes I and V using the above set-up. The frequency of observation was 
80 MHz
and the bandwidth was $\approx 1$ MHz. Both the Sun and the calibrator
sources can be treated as `point' sources since the fringe spacing or the first-null 
beam width (FNBW) of the interference pattern at the above frequency is 
broad ($\approx 5^{o}$). Figure \ref{fig9} shows the observations
of the circularly polarized radio emission from the Sun on 5 October 2012
with the above set-up. The emission is primarily due to the presence of a noise
storm source in the solar atmosphere which is known to be 
circularly polarized \citep{Elg77,Ram11a,Ram13}.
Contribution to the Stokes V emission from the `undisturbed' 
background solar corona is expected to be relatively small, 
particularly at 80 MHz \citep{Sas09}.
We would like to mention here that the effect of the 
instumental circular polarization in our observations can be
considered to be very small since the deflection in the Stokes V channel 
while observing some of the strong unpolarized calibrator sources was
less than the 3$\sigma$ level, where $\sigma$ is the rms noise in the system.

\section{Summary}

We have reported the design and performance of a CLPD in the frequency
range $\approx 68-84$ MHz with VSWR $<$ 2 and isolation $<$ -30 dBm 
between the two mutually orthogonal LPDs that constitute the CLPD. 
Trial observations indicate that the antennas 
can be used to effectively observe the circularly polarized radio
emission from the solar atmosphere with minimal instrumental
polarization. It is possible that the CLPD described can  
also be used to transmit and receive circularly polarized radio waves 
in applications involving transmission through the Earth's 
ionosphere which may
produce rotation of the wave polarization (particularly at
low frequencies). Note the polarization
arriving at the receiver from a linearly-polarized transmitting 
antenna may be practically unpredictable due to reasons mentioned
earlier. The compact size of the CLPD described also makes them suitable
for use as a primary antenna (feed) that illuminates a parabolic reflector
\citep{Smi01}.
We intend to extend the 
bandwidth of the CLPD and simulatenously observe radio emission from
different levels in the solar corona, with an array of CLPDs.

\section*{Acknowledgment}
It is a pleasure to thank the staff of the Gauribidanur observatory for their help in
the fabrication, testing of the antenna system and the observations.
\bibliographystyle{apj}
%\bibliographystyle{apj,clpd}
%\bibliographystyle{plain}
%\begin{thebibliography}{1}
\bibliography{clpd}

\end{document}